# Probing temperature-driven flow lines in a gated two-dimensional electron gas with tunable spin-splitting

Yi-Ting Wang[1], Gil-Ho Kim[2], C F Huang[1], Shun-Tsung Lo[3], Wei-Jen Chen[1], J T Nicholls[4], Li-Hung Lin[5], D A Ritchie[6], Y H Chang[1, 3], C-T Liang[1, 3], and B P Dolan[7, 8]

[1]*Department of Physics, National Taiwan University, Taipei 106, Taiwan*

[2]*School of Electronic and Electrical Engineering and Sungkyunkwan Advanced Institute of Nanotechnology (SAINT), Sungkyunkwan University, Suwon 440-746, Republic of Korea*

[3]*Graduate Institute of Applied Physics, National Taiwan University, Taipei 106, Taiwan*

[4]*Department of Physics, Royal Holloway, University of London, Egham, Surrey TW20 0EX, United Kingdom*

[5]*Department of Electrophysics, National Chiayi University, Chiayi 600, Taiwan*

[6]*Cavendish Laboratory, J. J. Thomson Avenue, Cambridge CB3 0HE, United Kingdom*

[7]*Department of Mathematical Physics, National University of Ireland, Maynooth, Ireland*

[8]*School of Theoretical Physics, Dublin Institute for Advanced Studies, 10 Burlington Road, Dublin, Ireland*

e-mail: ghkim@skku.edu, ctliang@phys.ntu.edu.tw and bdolan@thphys.nuim.ie

Abstract

We study the temperature flow of conductivities in a gated GaAs two-dimensional electron gas (2DEG) containing self-assembled InAs dots and compare the results with recent theoretical predictions. By changing the gate voltage, we are able to tune the 2DEG density and thus vary disorder and spin-splitting. Data for both the spin-resolved and spin-degenerate phase transitions are presented, the former collapsing to the latter with decreasing gate voltage and/or decreasing spin-splitting. The experimental results support a recent theory, based on modular symmetry, which predicts how the critical Hall conductivity varies with spin-splitting.

## I. Introduction

The magnetic-field-induced transitions observed in quantum Hall (QH) systems [1-8] are good examples of second-order quantum phase transitions [2, 3] described by scaling theory [4, 5]. By tracing the temperature ($T$)-dependence of the longitudinal and Hall conductivities $\sigma_{xx}$ and $\sigma_{xy}$, the $T$-driven flow diagram has been constructed for the transition regions separating adjacent QH phases with increasing perpendicular magnetic field $B$ [5-9]. In such a diagram, the unstable $T$-driven flow lines terminate at a critical point in the magnetic-field-induced transitions, and other flow lines tend toward the stable points corresponding to QH phases with decreasing $T$. A connection between the QH effect and the modular group was first suggested by Shapere, and Wilczek [10] and the correct sub-group of the modular group, appropriate for odd denominators in the fractional quantum Hall effect was identified by Lütken [11].

The modular group $\Gamma(1)$ is defined as the group of linear fractional transformations of the upper half complex plane which has the form

$$\sigma \rightarrow (a\sigma+b)/(c\sigma+d),$$

where $\sigma$ is a complex number, $a$, $b$, $c$, and $d$ are integers, and $ad-bc = 1$. Symmetric features such as the semicircle law [12] are often referred to the modular symmetry within the framework of the modular group. Interestingly, the semicircle law [12] and universal critical conductivities [2, 3] are expected in the temperature flow diagram based on the modular symmetry [6-9] built from Landau-level addition ($\sigma \rightarrow \sigma + 1$), flux attachment ($\sigma \rightarrow \sigma/(2\sigma+1)$), and particle-hole transformations ($\sigma \rightarrow 1 - \bar{\sigma}$) [1], where $\bar{\sigma}$ is the complex conjugate of $\sigma = \sigma_{xy} + i\sigma_{xx}$. The first two transformations can be taken as the generators of the group usually denoted by $\Gamma_0(2)$, a subgroup of the modular group $\Gamma(1)$ which is also important in elementary-particle physics [7], while the particle-hole transformation is an outer auto-morphism of $\Gamma_0(2)$.

When spin-splitting is fully resolved, the $T$-driven flow diagram following $\Gamma_0(2)$ modular symmetry has been constructed for the magnetic-field-induced transitions in the two-dimensional electron gases (2DEGs) in GaAs/AlGaAs heterostructures [2, 6]. The black dashed curves in Fig. 1 illustrate such a flow diagram. The flow lines approach the semicircles in the $\sigma_{xx}$-$\sigma_{xy}$ plane with decreasing $T$. In particular, the unstable flow lines converge toward the apexes of the semicircles at low temperatures, such that $\sigma_{xx}$ and $\sigma_{xy}$ become $T$-independent at the critical points indicated by the two black bullets. The modular symmetry, however, can be reduced to $\Gamma(2)$ when the resolved spin-splitting is small [6, 13-15]. The semicircle law remains valid under $\Gamma(2)$ symmetry, but the unstable critical points can deviate from the apexes of the semicircles [6, 15, 16]. When the spin-splitting is further reduced, the diameters of the semicircles should double for the spin-degenerate magnetic-field-induced transitions, whereas the $T$-driven flow diagram has features of $\Gamma_0(2)$ symmetry again, as indicated by the blue semicircle and blue bullet shown in Fig. 1 [15]. In this paper the spin-degenerate symmetry is denoted by $\Gamma^0(2)$ as described in Ref. [15]. The features of $\Gamma_0(2)$ and $\Gamma^0(2)$ have already been observed [6, 11-13], but more studies are necessary to clarify how the modular symmetry changes from $\Gamma_0(2)$ to $\Gamma(2)$ and $\Gamma^0(2)$ as the spin-splitting is varied.

The red arrow and bullet in Fig. 1 show how the two spin-resolved critical points corresponding to the same Landau-index $n$ merge to the single spin-degenerate one as the spin-splitting is suppressed [15]. Here $n$=0, 1, and 2,…correspond to the ground, 1st excited, 2nd excited Landau bands, and so on, respectively. In Fig. 1 we assume that the spin-splitting is fully resolved such that the two critical points are located at $(\sigma_{xy}, \sigma_{xx})=((4n+3)e^2/2h, e^2/2h)$ and $((4n+1)e^2/2h, e^2/2h)$, where $e$ and $h$ are the electron charge and Planck constant, respectively. With decreasing spin-splitting, the right critical point moves to the left along the spin-resolved semicircle

$$(\sigma_{xx}-e^2/2h)^2+[\sigma_{xy}-(4n+3)(e^2/2h)]^2=(e^2/2h)^2 , \qquad (1)$$

such that the corresponding critical Hall conductivity is reduced, although there can be bowing effects. The left critical point moves to the right along the other spin-resolved semicircle

$$(\sigma_{xx}-e^2/2h)^2+[\sigma_{xy}-(4n+1)(e^2/2h)]^2=(e^2/2h)^2 , \qquad (2)$$

becoming closer to the right one if bowing effects are negligible. With reduced spin-splitting, these two critical points merge to a single one as their critical Hall conductivities approach $(2n+1) e^2/h$. It should be noted that for $\Gamma^0(2)$ the flow lines do not approach the semicircle

$$(\sigma_{xx}-e^2/h)^2+(\sigma_{xy}-(2n+1)e^2/h)^2=(e^2/h)^2 \qquad (3)$$

when the spin-splitting collapses. In this case, the particle-hole symmetry remains valid and the unstable flow line is expected to lie on the vertical line $\sigma_{xy}=(2n+1) e^2/h$. Therefore the critical point can be located on the red bullet in Fig. 1. The critical point can move upward along $\sigma_{xy}=(2n+1)e^2/h$ with a further reduction of the spin-splitting. The semicircle law for $\Gamma^0(2)$ becomes valid after the critical point reaches the blue circle, which is located at the semicircle given by Eq. (3). To understand the effects of spin-splitting on the modular symmetry and temperature flow diagrams, we will study experimentally the traces of the critical points shown in Fig. 1. The predictions of modular symmetry are quite robust and varying the detailed assumptions still leads to similar flows [17].

To study the scaling theory under varying spin-splitting, we have re-analysed the data published in Ref. [18] to construct the $T$-driven flow lines in the $\sigma_{xx}$-$\sigma_{xy}$ plane. Previous studies show [18] that both types of magnetic-field-induced transitions, plateau-plateau (P-P) and insulator (I)-QH transitions, are observed at low $T$ as the perpendicular magnetic field $B$ is swept. In the following we denote each P-P transition by its adjacent QH states, for example, the transition separating the $\nu=1$ and $\nu=2$ QH states is denoted as a 2-1 or 1-2 transition. In addition, the 0-$\nu$ transition describes the I-QH transition where the 2DEG enters the $\nu$ QH state directly from the insulator designated by n=0. Decreasing the gate voltage reduces the spin-splitting by enhancing the strength of the disorder [19], and the following observations support the predictions described in Ref. [15] about spin effects on the modular symmetries:

(1) The spin-resolved and spin-degenerate transitions are governed by $\Gamma_0(2)$ and $\Gamma^0(2)$ modular symmetries, respectively.
(2) The unstable $T$-driven flow line is along $\sigma_{xy}=e^2/h$ when the 2-1 and 1-0 transition collapses into the 2-0 transition, consistent with particle-hole symmetry even when the semicircle law is invalid.
(3) The critical Hall conductivity in the 2-1 transition decreases continuously from $3e^2/2h$ to $e^2/h$ in the $\Gamma_0(2)$-$\Gamma(2)$-$\Gamma^0(2)$ crossover, in agreement with theory [15].
(4) There is no clear critical point in the 1-0 transition as the $\nu=1$ QH state is almost destroyed, although the 2DEG remains insulating for $\sigma_{xy}<e^2/2h$. Therefore, the critical Hall conductivity, if it exists, should become greater than $e^2/2h$ before such a transition is replaced by the 2-0 transition.

The device used in this study is a gated GaAs/AlGaAs heterostructure in which disorder is provided by InAs quantum dots and has been studied before [18]. In section II we focus on the case when the 0-2, 2-1 and 1-0 transitions are all observed at a certain gate voltage. The effects of varying the gate voltage and/or disorder on the 2-1 and 1-0 transitions are presented in section III and IV. In section V we investigate the temperature-driven flow lines in the 2-0 transition. Discussion and conclusions are made in section VI and VII, respectively.

**Section II Temperature flow diagrams following $\Gamma_0(2)$ and $\Gamma^0(2)$ modular symmetries**

At low magnetic fields, the sample behaves as an insulator [18], in the sense that the longitudinal resistivity decreases with decreasing temperature, $d\rho_{xx}/dT<0$. At $V_g=$ -0.264 V, there is no spin-resolved QH state for $B<1.4$ T and the sample undergoes a spin-degenerate 0-2 transition at the critical magnetic field $B_c^{0\text{-}2} = 1.13$ T, where $\rho_{xx}$ is $T$-independent. The expected universal value $h/2e^2$ indicates that we do not need to renormalise $\rho_{xx}$ for gated 2DEGs [6]. With increasing $B$, the spin-resolved QH state at $\nu=1$ appears and the 2DEG then undergoes a 1-0 transition at the critical field $B_c^{1-0}$ =3.42 T. In addition to the 0-2 and 1-0 transitions, the 2DEG undergoes a 2-1 transition. Figure 2 (a) – (c) show the constructed temperature flow diagrams for the 0-2, 2-1 and 1-0 transitions after the resistivities have been converted to conductivities using $\sigma_{xx}=\rho_{xx}/(\rho_{xx}^2+\rho_{xy}^2)$ and $\sigma_{xy}=\rho_{xy}/(\rho_{xx}^2+\rho_{xy}^2)$. In Fig. 2, each flow line is constructed by tracing ($\sigma_{xy}$, $\sigma_{xx}$) with respect to the temperature at a specific magnetic field, and the arrows indicate the flow direction with decreasing $T$.

In Figs. 2 (a) and (b), the vertical dotted lines denoting the $T$-driven flow lines at the critical magnetic fields $B_c^{0-2}$ and $B_c^{1-0}$ are the unstable ones. Figure 2 (b) shows that the flow lines to the right and left of the apex flow toward the expected stable points at ($e^2/h$,0) and (0,0) along the semicircle described by Eq. (2) with $n=0$. In addition, there is unstable flow close to the apex at ($\sigma_{xy}$, $\sigma_{xx}$)=($e^2/2h$, $e^2/2h$). Therefore, the temperature flow diagram for the 0-1 transition has the characteristics of $\Gamma_0(2)$ symmetry. Figure 3 (a) shows that the curves of $\sigma_{xy}$ at different temperatures intersect at the critical point, which is indicated by the dotted line, because $\sigma_{xy}$ is $T$-independent at such a point as expected. Figure 2 (a) shows that the $T$-driven flow lines on the right hand side of the unstable one are along the semicircle described by Eq. (3) with $n=0$. Therefore, the 0-2 transition can be assigned $\Gamma^0(2)$ rather than $\Gamma_0(2)$ modular symmetry, since there is no spin-resolved state at low $B$.

The expected semicircle for the 2-1 transition is given by Eq. (1) with $n=0$, where the apex has critical longitudinal and Hall conductivities of $\sigma_{xx}^c = e^2/2h$ and $\sigma_{xy}^c$ =$3e^2/2h$. Figure 3 (b) shows that $\sigma_{xy}$ is $T$-independent at the point $B_c^{2-1}$ when $\sigma_{xy}$~1.5 $e^2/h$, which is expected under $\Gamma_0(2)$ symmetry. In Fig. 2 (c), the unstable flow occurring for $\sigma_{xy}$~1.5 $e^2/h$ does not converge to the apex of the semicircle for the 2-1 transition, since over this range $\sigma_{xx}$ is not $T$-independent and is smaller than 0.5 $e^2/h$. We note that $\sigma_{xy}$ and $\sigma_{xx}$ correspond to the coefficients of the topological and kinetic terms, respectively, where the first term is important in constructing the phase diagram of the quantum Hall effect [13]. As reported previously [13], the temperature flow features in $\sigma_{xy}$ can be more robust than those in $\sigma_{xx}$, and the critical Hall conductivity indicates the $\Gamma_0(2)$ symmetry in the 2-1 transition. Therefore it is useful to study $\sigma_{xy}(T)$ at different magnetic fields, and figure 4 (a) shows the $T$-independent behaviour (dotted) of $\sigma_{xy}$ at $B_c^{2-1}$ separating the two regions in the $T$-$\sigma_{xy}$ plane expected from scaling theory: the lower region corresponds to the $\nu$=1 QH state, whereas the upper one corresponds to the $\nu$=2 QH state. For $T$<0.45 K, $\sigma_{xy}$ increases in the upper region and reaches the quantized value $2e^2/h$, whereas $\sigma_{xy}$ decreases in the lower region and approaches $e^2/h$ as $T\rightarrow 0$. At $V_g$= -0.264 V, $\Gamma_0(2)$ symmetry can be identified in the spin-resolved 2-1 and 1-0 transitions, whereas the temperature flow diagram for the spin-degenerate 0-2 transition has features of $\Gamma^0(2)$ symmetry.

**III The critical Hall conductivity of the 2-1 transition at different gate voltages**

At $V_g$ = -0.264 V, see Fig. 4 (a), the value of the critical Hall conductivity suggests that the 2-1 transition exhibits the $\Gamma_0(2)$ symmetry. So at low temperatures $\sigma_{xy}$ converges to ~$3e^2/2h$, between the two quantized conductivities for the $\nu$=1 and 2 QH states. Decreasing the gate voltage to -0.280 V increases the disorder [18], and the 2-1 and 1-0 transitions merge to the 2-0 transition. Figure 4 (b) shows that at $V_g$=-0.280 V $\sigma_{xy}$ is $T$-independent with a value of $e^2/h$, the expected critical value for the 0-2 transition under $\Gamma^0(2)$ symmetry,.

Figure 4 (c)-(e) show the $T$-dependence of $\sigma_{xy}$ for -0.276 V $\leq V_g \leq$ -0.272 V. In these diagrams, we investigate the 2-1 transition before it merges with the 1-0 transition. At low temperatures, the $T$-independent curves (indicated by the horizontal dot lines) appear at $\sigma_{xy}$ ~ 1.45, 1.33, and 1.23 $e^2/h$ when the gate voltage is decreased from -0.272 V, -0.274 V, to -0.276 V, respectively. Therefore figures 4 (a) to (e) show that the critical Hall conductivity $\sigma_{xy}^c$ of the 2-1 transition decreases from $3e^2/2h$ toward $e^2/h$ and then the 2-1 transition merges with the 1-0 transition. The critical values $3e^2/2h$ and $e^2/h$ are the expected ones for the spin-resolved 2-1 and spin-degenerate 2-0 transitions under $\Gamma_0(2)$ and $\Gamma^0(2)$ symmetries, respectively, and the continuous decrease of $\sigma_{xy}^c$ supports the prediction [15] on the appearance of reduced modular symmetry $\Gamma(2)$.

The $\nu$=1 QH state is due to spin-splitting, which is much smaller than the cyclotron gap that produces the $\nu$=2 QH state in GaAs-based 2DEGs. With decreasing gate voltage (increasing disorder), it is reasonable to expect that the spin gap will be destroyed before the cyclotron gap. From the minima of $\rho_{xx}$ ($\sigma_{xx}$) in the $\nu$=1 QH state, we can see that the spin-splitting is reduced under the increase of the disorder.

**IV The critical Hall conductivity in the 1-0 transition**

At $V_g$ = -0.280 V, there is no $\nu$=1 QH state and the 2DEG undergoes a 2-0 transition to become a high-field insulator. Figure 4 (b) shows that for $0 < \sigma_{xy} < e^2/h$, $\sigma_{xy}$ decreases with decreasing $T$ since the 2DEG is an insulating phase. At $V_g$ = -0.264V, a $\nu$=1 QH state exists and the 2DEG undergoes 1-0 transition rather than the 2-0 transition. Figure 5 (a) shows that at $V_g$ = -0.264 V $\sigma_{xy}$ is $T$-independent near 0.5 $e^2/h$, the expected critical value under $\Gamma_0(2)$ symmetry. For the 2-1 transition, $\sigma_{xy}^c$ decreases from ~1.5 $e^2/h$ to 1.33 $e^2/h$ as $V_g$ is decreased from -0.264 V to -0.274 V as shown in Figs. 5 (a)-(c). In contrast, for the 1-0 transition the corresponding $\sigma_{xy}^c$ =(0.5±0.05) $e^2/h$ and thus remains ~0.5 $e^2/h$. Therefore the 1-0 transition satisfies $\Gamma_0(2)$ symmetry, whereas the reduced modular symmetry $\Gamma(2)$ should be taken into account in the 2-1 transition. In our system, the 1-0 transition is observe at a higher magnetic field than that for the 2-1 transition. Therefore the spin-splitting for the 1-0 transition can be larger than that for the 2-1 transition. In this case, the 1-0 transition can be described by $\Gamma_0(2)$ symmetry, and there is no reduction on the modular symmetry.

In our study, the 1-0 transition occurs at $V_g$ = -0.274 V, and merges with the 2-1 transition to become a 2-0 transition at $V_g$ = -0.280 V. Figure 5 (d) shows that $\sigma_{xy}$ is almost $T$-independent for (0.5-0.9) $e^2/h$ at $V_g$ = -0.276 V. We note that a quantum phase transition is defined as $T$ approaches zero, whereas experiments are always performed at finite temperatures. Therefore finite temperature effects should be considered in a real system. Although $\Gamma(2)$ symmetry can be identified in the 2-1 transition at $V_g$ = -0.276 V, there could be competition between $\Gamma(2)$ and $\Gamma_0(2)$ symmetries in the 1-0 transition because the spin-splitting is larger for $\Gamma_0(2)$. If there was competition, there would be no clear transition point at finite temperatures and thus the Hall conductivity would be $T$-independent over a wide range of magnetic field at $V_g$ = -0.276 V. The bowing effects [13] resulting from $\Gamma(2)$ symmetry could also exist in the competition. Figure 5 (a)-(c) shows that $\sigma_{xy}^c$ can be slightly below $e^2/2h$. At $V_g$=-0.276 V, $\sigma_{xy}$ decreases with decreasing $T$ when $\sigma_{xy}$ becomes smaller than 0.5 $e^2/h$, and hence the $T$-independent critical point, if it exists, will not decrease to become smaller than $e^2/2h$. So the critical Hall conductivity should have the tendency to increase from ~$e^2/2h$ to $e^2/h$ as predicted in Ref. [15].

**V. The temperature-driven flow lines in the 2-0 transition**

In addition to the 2-1 and 1-0 transitions, we have observed a 0-2 transition at lower $B$ at $V_g$ = -0.264 V. When the gate voltage is decreased, the spin-splitting is not observed and thus the corresponding $T$-driven flow lines show $\Gamma^0(2)$ symmetry as the 2DES enters $\nu$=2 QH state from the insulator at low $B$. Figure 6 shows the $T$-driven flow diagram at $V_g$=-0.280 V. the flow lines in the 0-2 transition are close to the upper semicircle defined by Eq. (3) with $n=0$, and the directions of flow are opposite for $\sigma_{xy} < e^2/h$ and $\sigma_{xy} > e^2/h$. Therefore $\Gamma^0(2)$ symmetry is valid for the case at $V_g = -0.280$ V. In addition, the unstable point is located near the apex $(\sigma_{xy}, \sigma_{xx}) = (e^2/h, e^2/h)$ of this semicircle. So the critical longitudinal and Hall conductivities are of the expected universal values at low $B$, whereas the critical points at high $B$ depend on the spin-splitting.

It should be noted that at $V_g = -0.280$ V, there is an additional 2-0 transition due to the collapse of the 2-1 and 1-0 transitions. It is predicted [15] that $\sigma_{xx} < e^2/h$ and the critical point is expected to leave the semi-circle as the collapse occurs. Unstable flow is expected along $\sigma_{xy} = e^2/h$ due to particle-hole symmetry for $\Gamma(2)$ and $\Gamma^0(2)$. Figure 6 shows the flow lines in the transition resulting from the collapse are not along any semi-circle (although semi-circles are still flow lines, the critical point no longer lies on these semi-circles after the collapse). Moreover, the flow line along $\sigma_{xy} = e^2/h$ is the unstable one because the flow directions are opposite for $\sigma_{xy} < e^2/h$ and $\sigma_{xy} > e^2/h$. Furthermore, the longitudinal conductivity $\sigma_{xx}$ is smaller than $e^2/h$ along the unstable flow. So our experimental study supports the prediction on particle-hole symmetry, even when the semicircle law is not valid [15].

As a result of the particle-hole symmetry, the flow lines in the 2-0 transition are expected to be symmetric with respect to the vertical line $\sigma_{xy} = e^2/h$ in the $\sigma_{xy}$-$\sigma_{xx}$ plane [6, 13, 15]. In contrast, figure 6 shows that the flow lines for the 2-0 transition are asymmetric with respect to $\sigma_{xy} = e^2/h$; this asymmetry could be caused by the spin-splitting which varies with gate voltage. A full description probably requires a three-dimensional flow, with Zeeman energy as a third axis. The particle-hole symmetry, in fact, is important to the appearance of the unstable flow along $\sigma_{xy} = e^2/h$. We note that in Ref. [19], the flow diagram symmetric to $\sigma_{xy} = e^2/h$ can be obtained when the $\nu=1$ QH state appears with decreasing $T$.

## VI. Discussion

It was proposed that the collapse of spin-splitting in the integer quantum Hall effect is due to disorder broadening of Landau levels [20], and this is what we observe. In our system, with more negative gate voltage (increased disorder), the spin-splitting collapses near $-0.280$ V $\leqq V_g \leqq -0.276$ V and the $\nu=1$ QH state disappears. Therefore, for $V_g \leqq -0.276$ V we observe a 2-0 transition instead of a 2-1-0 transition, in agreement with the Folger-Shklovskii model [20].

When the spin-splitting is fully resolved, see Fig. 1, the two critical points are located at $(3e^2/2h, e^2/2h)$ and $(e^2/2h, e^2/2h)$ for $n=0$. These two points approach the point $(e^2/h, 0)$, which corresponds to the $\nu=1$ QH state, along the semicircles described by Eqs. (1) and (2) as the spin-splitting is reduced but still can be resolved. The QH state occurs between the two critical points when the Fermi energy is in the localized states, and the width of the QH state in $B$ is determined by the distance between the two critical points. Therefore a poorly developed $\nu=1$ QH state is expected at finite temperatures when the critical points are close to ($e^2/h$, 0). To probe the poorly defined QH state, Figs. 7 (a) and (b) show the curves of $\rho_{xx}$ and $\rho_{xy}$ for $V_g =$ -0.264 V and $T$=0.21 K and $V_g =$ -0.274 V and $T$=0.18 K, respectively. In Fig. 7 (a), there is a peak in $\rho_{xx}$ at $B\sim 2$ T separating the $\nu=2$ and 1 QH states characterized by their quantized Hall plateaus. In addition, $\Gamma^0(2)$ symmetry is observed in both the 2-1 and 1-0 transitions. With decreasing disorder, the decrease of the critical Hall conductivity of the 2-1 transition shows that the modular symmetry is reduced to $\Gamma(2)$ when -0.274 V < $V_g$ < -0.268 V. Figure 7 (b) shows that at $V_g =$ -0.274 V there is no peak in $\rho_{xx}$ separating the $\nu=2$ and 1 QH states, even though the $\nu=2$ and $\nu=1$ plateaus can be observed in $\rho_{xy}$. We note that the position of the minimum in $\rho_{xx}$ at $\nu=2$ in $B$ is slightly higher than that of the mid-point of the QH $\nu=2$ plateau. Such a result is not caused by a carrier density inhomogeneity in our sample. The reason for this is that, as shown in Fig. 8, we have observed well-defined temperature-independent points in $\rho_{xx}$ which correspond to insulator-quantum Hall transitions at various gate voltages. Such data cannot be obtained in an inhomogeneous sample since for different carrier densities, there should be different crossing points in $\rho_{xx}$. Moreover, Fig. 8 shows that there is only a single period of the observed Shubnikov-de Haas-type oscillations over the whole measurement range. All these results demonstrate that the carrier density in our device is uniform. A possible reason for the difference between the $\rho_{xx}$ minimum and the centre of the QH $\nu=2$ plateau are electron-electron interactions, which are known to cause a correction to the Hall resistivity and to increase the Hall slope with decreasing temperature [21]. This effect is shown in Fig. 8, which shifts the centre of the $\nu=2$ QH plateau to a lower $B$, causing it to not coincide with the $\rho_{xx}$ minimum in $B$.

We note that the quantum Hall effect in $\rho_{xy}$ can be observed without there being a zero in $\rho_{xx}$ [22-26], and Fig. 7 (b) shows that it may become difficult to identify the poorly-defined QH state in $\rho_{xx}$. Although the $\nu=1$ QH state should correspond to the $T$-independent point at $\sigma_{xy}=e^2/h$, such a point can deviate from the expected value when this QH state is not well-defined [27]. Moreover, the weak $T$-dependence of $\sigma_{xy}$, see Fig. 5 (d), shows that $\sigma_{xy}$ can be $T$-independent over a wide range in $B$ although discrete $T$-independent points are expected. The collapse of the 2-1 and 1-0 transitions takes place when the two critical points merge, but it is not clear why this occurs. Nevertheless this study reveals that the $T$-independent Hall conductivity is observed over a wide range of $B$, rather than at a single transition point, under competition between $\Gamma^0(2)$ and $\Gamma(2)$ symmetries. We note that the critical points are expected to show an abrupt change on the vertical line $\sigma_{xy}=e^2/h$ under the collapse of the 2-1 and 1-0 transitions [15]. This study shows that the unstable flow is along $\sigma_{xy}=e^2/h$, when the 2-0 transition is replaced by the 2-1 and 1-0 transitions.

## VII. Conclusion

We have studied temperature flow of a highly-disordered gated two-dimensional electron gas with tunable disorder and/or spin-splitting. At $V_g = -0.264$ V, we observe a 0-2-1-0 transition. At low magnetic fields, the 0-2 transition is consistent with $\Gamma^0(2)$ symmetry. With increasing magnetic field, the 0-1 transition is compatible with $\Gamma_0(2)$ symmetry. With increasing disorder, it appears that the 2-1 transition changes from $\Gamma_0(2)$ to $\Gamma(2)$ symmetry when the spin gap becomes smaller. In addition, our experimental results show that the critical point in $\sigma_{xy}$ moves from $3e^2/2h$ toward $e^2/h$.

We note that the case for the 0-1 transition is not clear. It appears that the critical point or unstable flow stays at around $\sigma_{xy}=e^2/2h$ with decreasing spin gap. A possible reason is that the 0-1 transition occurs at a higher magnetic field compared to the 2-1 transition. Therefore the mixing of spin and disorder may be less significant for the 1-0 transition compared with the 2-1 transition. We note that the unstable flow shows an abrupt change to $\sigma_{xy} = e^2/h$ with $\sigma_{xx} < e^2/h$ when the 1-0 transition merges with the 2-1 transition, which supports the prediction based on the modular symmetry $\Gamma(2)$, with particle-hole symmetry, for the unstable flow when the spin-splitting becomes unresolved [15]. In addition, the $T$-dependence of $\sigma_{xy}$ at $V_g = -0.276$ V indicates that the critical point will not occur for $\sigma_{xy} <0.5\ e^2/h$ and thus should be larger than 0.5 $e^2/h$, if it exists, before the 1-0 transition to be replaced by the 2-0 transition. Therefore, our observation supports the prediction on the critical Hall conductivity with varying spin-splitting.

## Acknowledgments

This work was partly supported by National Taiwan University (grant number: 10R809083B). Gil-Ho Kim was supported by World Class University program funded by the Ministry of Education, Science and Technology through the National Research Foundation of Korea (R32-10204). C.T.L., Y.H.C. and L.H.L. acknowledge support from the NSC, Taiwan. D.A.R. acknowledges support from the EPSRC, UK.

Academic/Plenum Publisher, New York, USA

Figure Captions

Figure 1 Modular symmetry and temperature ($T$)-driven flow diagram in the longitudinal conductivity $\sigma_{xx}$ and Hall conductivity $\sigma_{xy}$ plane. For a spin-degenerate (spin-split) system, the semicircle curve in blue and the solid blue bullet point correspond (two semicircle curves in black and the solid black bullets) to the theoretically predicted $\sigma_{xy}(\sigma_{xx})$ at a low $T$ and the unstable point. With decreasing spin-splitting, it is predicted [15] that both of the black bullets move toward ($\sigma_{xy}$, $\sigma_{xx}$)=(($2n+1)e^2/h$, 0), as indicated by the black arrows. With reduced spin-splitting, the black bullets merge into the red bullet and move along the line $\sigma_{xy} = (2n+1)e^2/h$, as indicated by the red arrow. When the system becomes spin-degenerate, the unstable point is located at the apex of the blue semicircle curve.

Figure 2 $T$-driven flow lines of the (a) 0-2 transition, (b) 1-0 transition, and (c) 2-1 transition regions for $V_g$ = -0.264 V. The dashed curves indicate the theoretically predicted semicircles, and the arrows indicate the flow lines to the low measurement temperatures at a fixed magnetic field. The black bullets indicate that with decreasing temperature, an approximately $T$-independent $\sigma_{xy}$ is observed. The dotted line serve as a guide to the eye showing that $\sigma_{xy}$ is $T$-independent close to the predicted values of (a) $e^2/2h$, (b) $e^2/h$, and (c) $3e^2/2h$, respectively.

Figure 3 The conductivities $\sigma_{xy}(B)$ at various $T$ for $V_g$ = -0.264 V in (a) the 1-0 transition and (b) the 2-1 transition. The dotted lines indicate that temperature-independent $\sigma_{xy}$ are observed at the expected values of (a) $e^2/2h$ and (b) $3e^2/2h$, respectively.

Figure 4 The conductivities $\sigma_{xy}$ for a fixed $B$ at various $T$ for (a) $V_g$ = -0.264 V, (b) $V_g$ = -0.280 V, (c) $V_g$ = -0.272 V, (d) $V_g$ = -0.274 V, and (e) $V_g$ = -0.276 V, in the 2-1 transition region. The solid lines join the data points in $\sigma_{xy}$. The dotted lines are guides to the eye, indicating that $\sigma_{xy}$ is almost $T$-independent at low temperatures.

Figure 5 The conductivity $\sigma_{xy}$ for a fixed $B$ at various $T$ for (a) $V_g$ = -0.264 V, (b) $V_g$ = -0.272 V, (c) $V_g$ = -0.274 V, and (d) $V_g$ = -0.276 V in the 1-0 transition region. The solid lines join the data points in $\sigma_{xy}$. The dotted lines are guides to the eye, indicating that $\sigma_{xy}$ is almost $T$-independent at low temperatures.

Figure 6 The red curves correspond to $\sigma_{xy}(\sigma_{xx})$ at the lowest temperature of 0.03 K. The dashed curves and line are the expected semicircles and vertical one along

$\sigma_{xy}=e^2/h$, respectively. The different colours denote the temperature flow at a specific magnetic field from the highest to lowest temperature, as indicated by the arrows. The black bullet at the apex corresponds to the critical unstable point in the low-field 0-2 transition. The red bullets on the straight line $\sigma_{xy} = e^2/h$ correspond to case when the 2-1 and 1-0 transitions merge into the 2-0 transition.

Figure 7 (a) Longitudinal and Hall resistivities $\rho_{xx}(B)$ and $\rho_{xy}(B)$ as a function of magnetic field for $V_g$ = -0.264 V and $T$=0.21 K. There is a peak in $\rho_{xx}$ for $B$~2 T separating the $\nu$=2 and $\nu$=1 QH states characterised by the quantized Hall plateaus. (b) $\rho_{xx}(B)$ and $\rho_{xy}(B)$ for $V_g$ = -0.274 V and $T$=0.18 K. There is no peak in $\rho_{xx}$ separating the $\nu$=2 and $\nu$=1 QH states, even though the $\nu$=2 and $\nu$=1 plateaus can be seen in $\rho_{xy}$.

Figure 8 Longitudinal and Hall resistivities $\rho_{xx}(B)$ and $\rho_{xy}(B)$ as a function of magnetic field for $V_g$ = -0.264 V at various $T$.

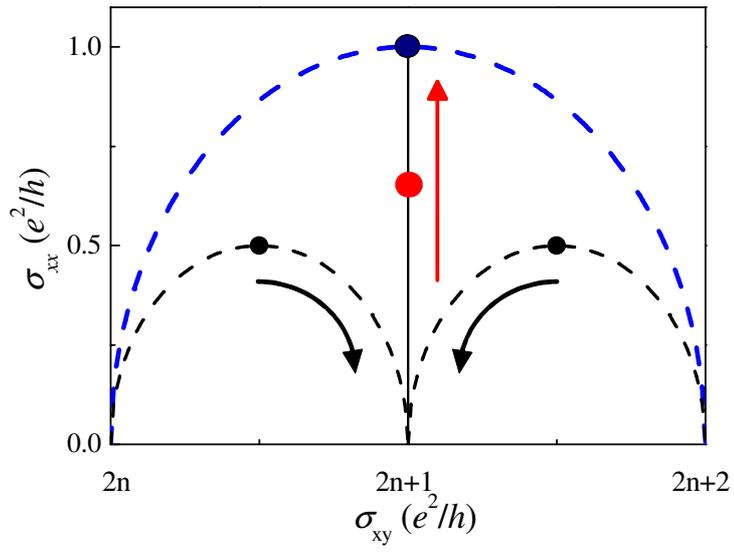

Fig. 1

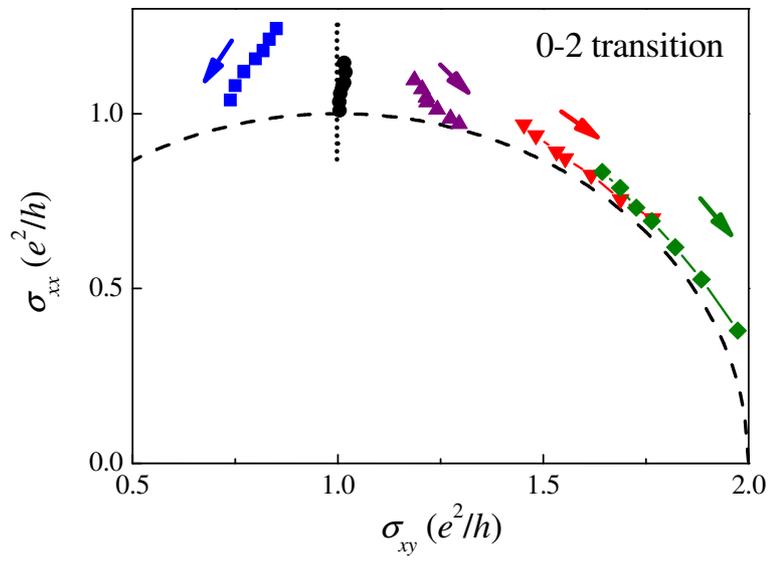

Fig. 2 (a)

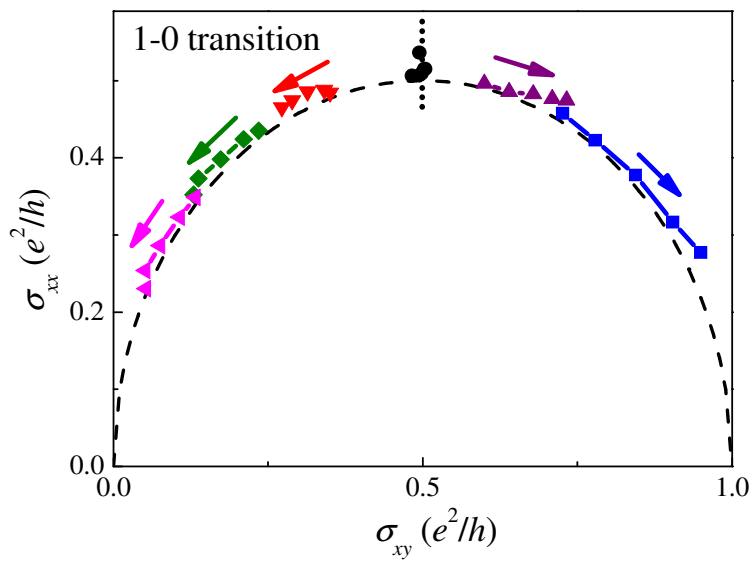

Fig. 2 (b)

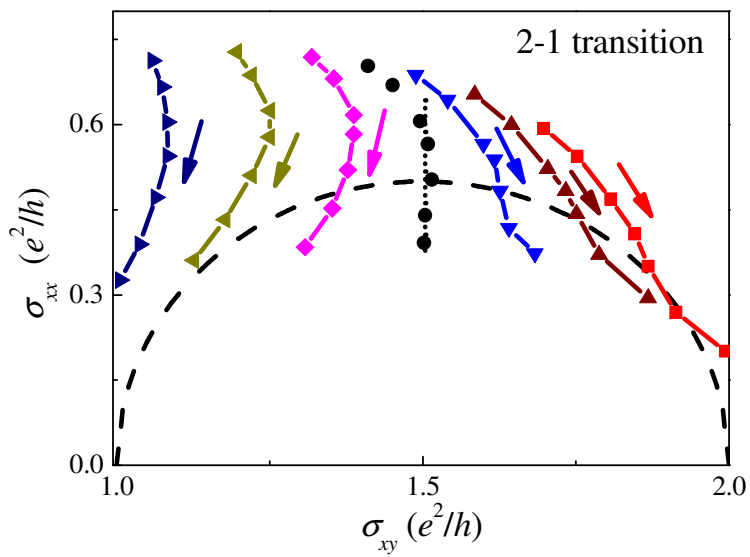

Fig. 2 (c)

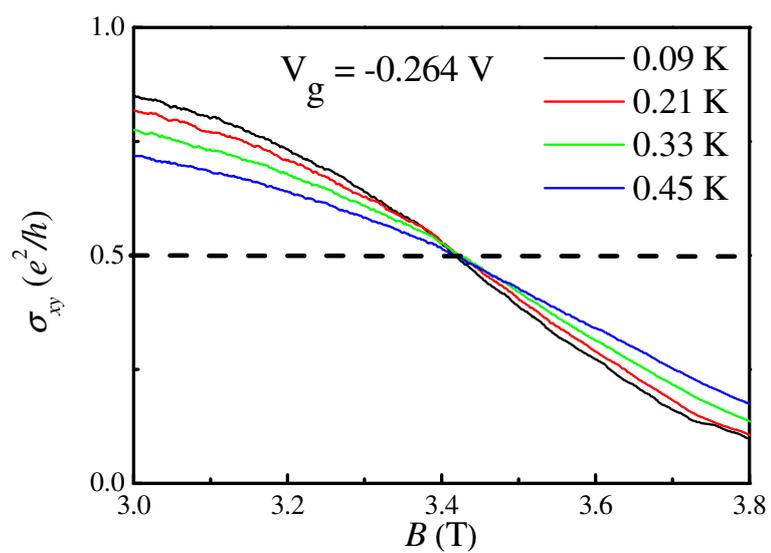

Fig. 3 (a)

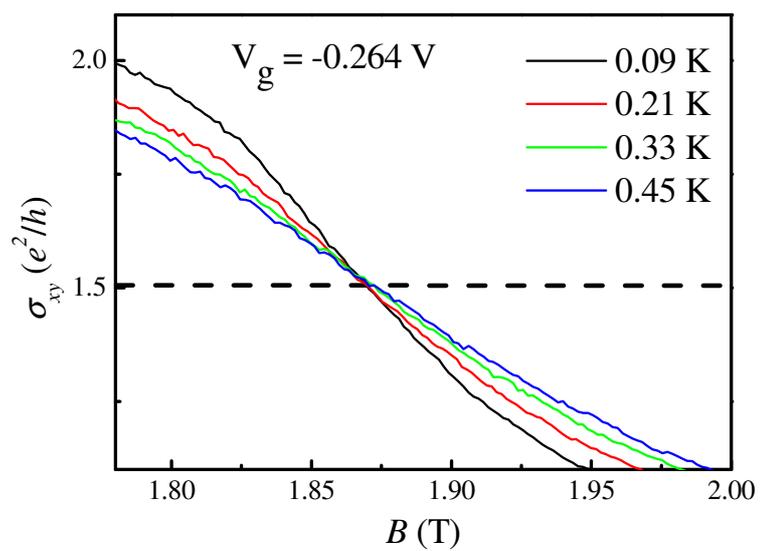

Fig 3 (b)

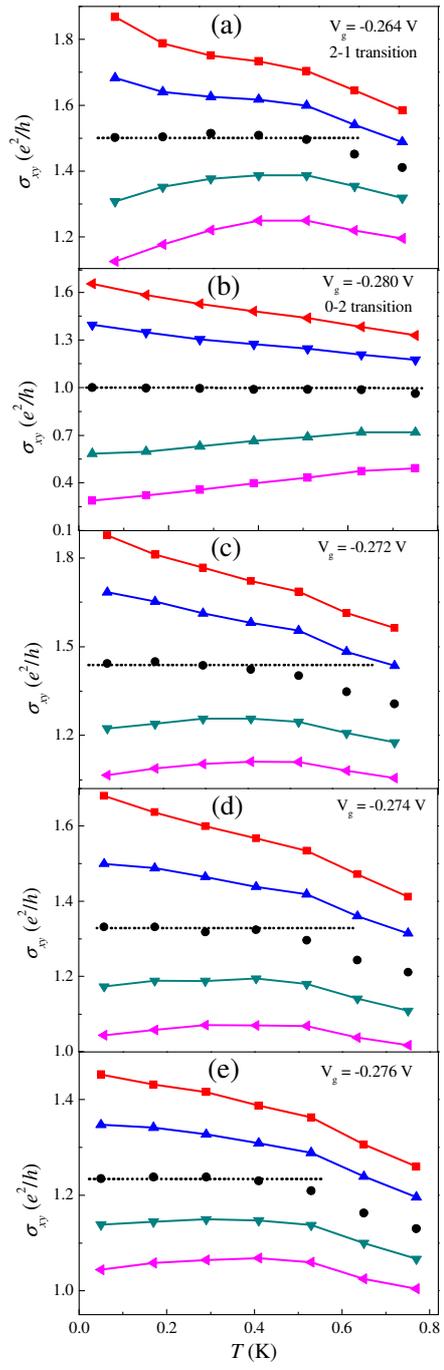

Fig. 4

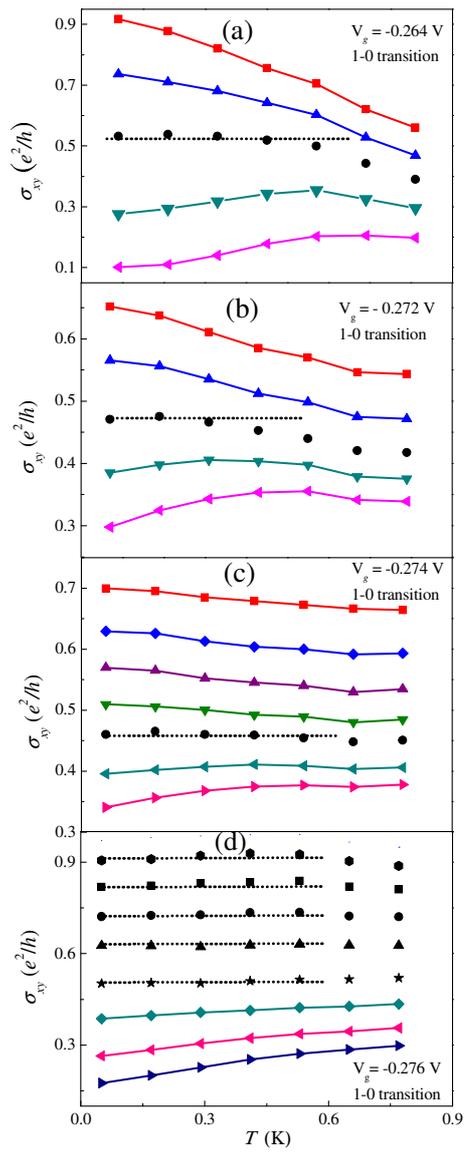

Fig. 5

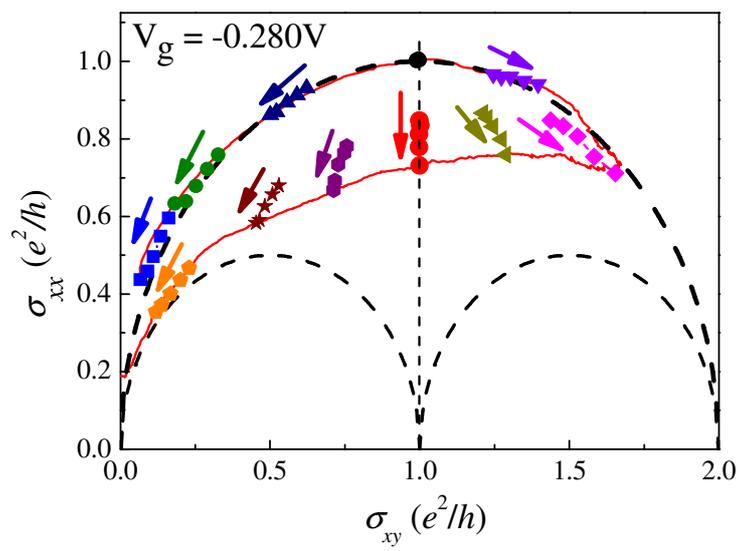

Fig. 6

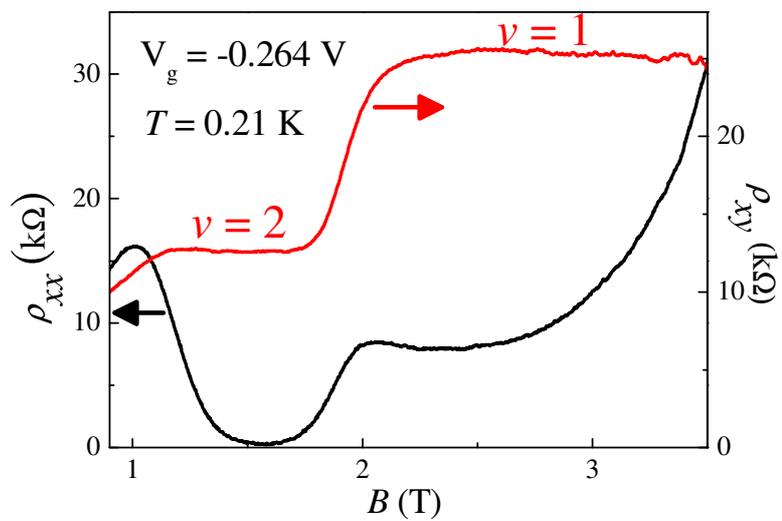

Fig. 7 (a)

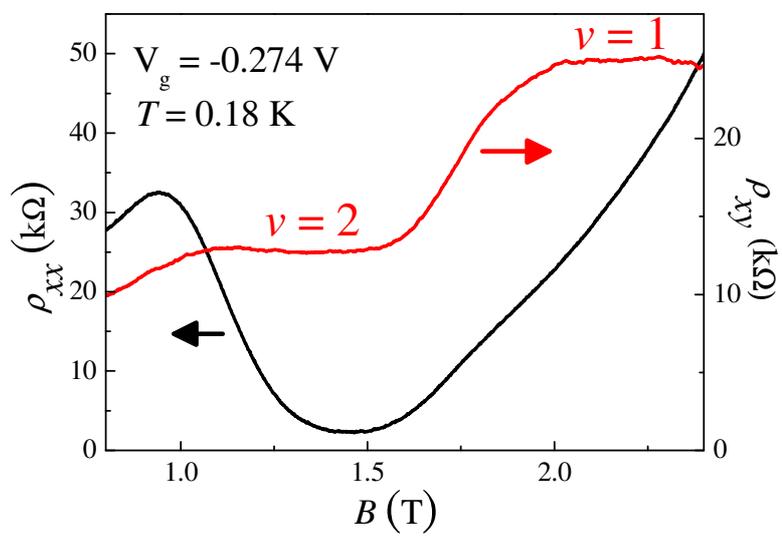

Fig. 7 (b)

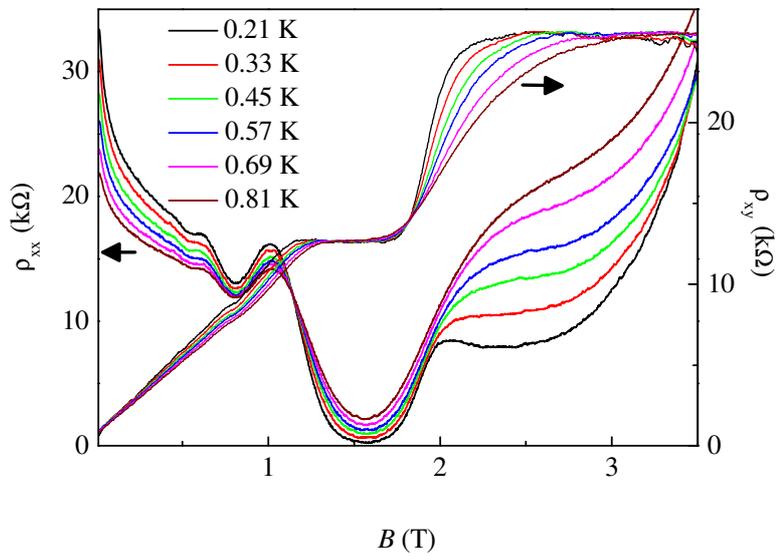

Fig. 8